\documentstyle[twoside,fleqn,npb,epsfig]{article}
\newcommand{\lamps}{\lambda_p^{(s)}}

\newcommand{\AmS}{{\protect\the\textfont2
  A\kern-.1667em\lower.5ex\hbox{M}\kern-.125emS}}

\title{Radiative corrections to radiative B decays:
        Exclusive $B\to V\gamma$ at NLO}

\author{G. Buchalla\address{Ludwig-Maximilians-Universit\"at M\"unchen,
                    Sektion Physik, \\
        Theresienstra\ss e 37, D-80333 M\"unchen, Germany}}%

\begin{document}

\begin{abstract}
We discuss a model-independent framework for the analysis
of the radiative $B$-meson decays $B\to K^*\gamma$ and $B\to\rho\gamma$
based on the heavy-quark
limit of QCD. We present a factorization formula for the
treatment of $B\to V\gamma$ matrix elements involving charm (or up-quark)
loops, which contribute at leading power in $\Lambda_{QCD}/m_B$ to
the decay amplitude. Annihilation topologies are power suppressed,
but still calculable in some cases.
The framework of QCD factorization is necessary to compute 
{\it exclusive\/} $b\to s(d) \gamma$ decays systematically beyond
the leading logarithmic approximation. 
Results to next-to-leading order in QCD
and to leading order in the heavy-quark limit are given
and phenomenological implications are discussed.
\end{abstract}

\maketitle

\section{INTRODUCTION}

The radiative transitions $b\to s\gamma$, $b\to d\gamma$ are among
the most valuable probes of flavour physics.
Among the characteristics are the high sensitivity to New Physics
and the particularly large impact of short-distance
QCD corrections.
Considerable efforts have therefore been devoted to achieve a full
calculation of the inclusive decay $b\to s\gamma$ at next-to-leading
order (NLO) in renormalization group (RG) improved perturbation
theory (see the talk by M. Misiak in these
proceedings).

Whereas the inclusive mode can be computed perturbatively, using
the fact that the $b$-quark mass is large and employing the heavy-quark
expansion, the treatment of the exclusive channel $B\to K^*\gamma$
is in general more complicated. In this case bound state effects
are essential and need to be described by nonperturbative 
hadronic quantities (form factors).
The basic mechanisms at next-to-leading order were already
discussed previously for the $B\to V\gamma$ amplitudes 
\cite{AAWGSW,DLTES}.
However, hadronic models were used to evaluate the various
contributions, which did not allow a clear separation
of short- and long-distance dynamics and a clean distinction
of model-dependent and model-independent features.

In this talk\footnote{
    At 6th International Symposium on Radiative Corrections
   (RADCOR) and 6th Zeuthen Workshop on Elementary Particle Theory
    (Loops and Legs), 8-13 September 2002, Kloster Banz, Germany;
     LMU 02/15}
 we present a systematic analysis of the exclusive
radiative decays $B\to V\gamma$ ($V=K^*$, $\rho$) in QCD, based
on the heavy quark limit $m_b\gg\Lambda_{QCD}$ \cite{BB,BFS}
(see also \cite{AP}). We shall provide
factorization formulas for the evaluation of the relevant hadronic
matrix elements of local operators in the weak Hamiltonian.
Factorization holds in QCD to leading power in the heavy quark limit.
This result relies on arguments similar to those used previously
to demonstrate QCD factorization for hadronic two-body modes
of the type $B\to\pi\pi$ \cite{BBNS}.

Within this approach higher order QCD corrections can be consistently
taken into account. We give the $B\to V\gamma$ decay amplitudes
at next-to-leading order (NLO).
After including NLO corrections the largest uncertainties still come
from the $B\to V$ form factors, which are at present known only with 
limited precision ($\sim \pm 15\%$), mostly from QCD
sum rule calculations \cite{BB98}, which we used in the
present analysis.

\section{BASIC FORMULAS}\label{sec:basics}

The effective Hamiltonian for $b\to s\gamma$ transitions reads
(see e.g. \cite{BB})
\begin{equation}\label{heff}
{\cal H}_{eff}=\frac{G_F}{\sqrt{2}}\sum_{p=u,c}\lamps
\left[\sum_{i=1,2} C_i Q^p_i +\sum^8_{i=3} C_i Q_i\right]
\end{equation}
where $\lamps=V^*_{ps}V_{pb}$
The four-quark operators $Q^p_{1,2}$ and the magnetic penguin
operator $Q_7$ give the most important contribution. 
The most difficult step in computing the decay amplitudes is the
evaluation of the hadronic matrix elements of the operators in 
(\ref{heff}). We will argue that in this case the following 
factorization formula is valid 
\begin{eqnarray}\label{fform}
&&\langle V\gamma(\epsilon)|Q_i|\bar B\rangle =
\Bigl[ F^{B\to V}(0)\, T^I_{i} + \nonumber \\
&&\ \ +\int^1_0 d\xi\, dv\, T^{II}_i(\xi,v)\, \Phi_B(\xi)\, \Phi_V(v)\Bigr]
\cdot\epsilon
\end{eqnarray}
where $\epsilon$ is the photon polarization 4-vector.
Here $F^{B\to V}$ is a $B\to V$ transition form factor,
and $\Phi_B$, $\Phi_V$ are leading twist light-cone distribution amplitudes
(LCDA) of the $B$ meson and the vector meson $V$, respectively.
These quantities 
describe the long-distance dynamics of the matrix elements, which
is factorized from the perturbative, short-distance interactions
expressed in the hard-scattering kernels $T^I_{i}$ and $T^{II}_i$.
The QCD factorization formula (\ref{fform}) holds up to
corrections of relative order $\Lambda_{QCD}/m_b$.

\begin{figure}[t]
   \epsfysize=2.5cm
   \epsfxsize=4.5cm
   \centerline{\epsffile{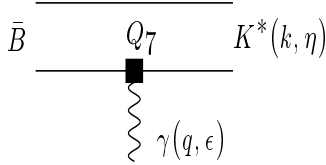}}
   \vspace*{-1cm}
\caption{Contribution of the magnetic penguin operator $Q_7$
described by $B\to V$ form factors. \label{fig:q7}}
\end{figure}

For $Q_7$ the factorization formula (\ref{fform}) is trivial.
The matrix element is simply expressed in terms of the standard
form factor, $T^I_{7}$ is a purely kinematical function and
the spectator term $T^{II}_7$ is absent. An illustration is given
in Fig. \ref{fig:q7}.
In the leading logarithmic approximation
(LO) and to leading power in the heavy-quark limit, $Q_7$ gives the
only contribution to the amplitude of $\bar B\to V\gamma$.

\begin{figure}[t]
   \epsfysize=2.5cm
   \epsfxsize=8cm
   \centerline{\epsffile{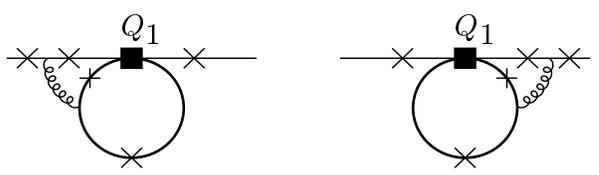}}
   \vspace*{-1cm}
\caption{${\cal O}(\alpha_s)$ contribution to the hard-scattering 
kernels $T^I_{i}$ from four-quark operators $Q_i$.
The crosses indicate the places where the emitted photon can
be attached. \label{fig:qit1}}
\end{figure}

\begin{figure}[t]
   \epsfysize=2cm
   \epsfxsize=8cm
   \centerline{\epsffile{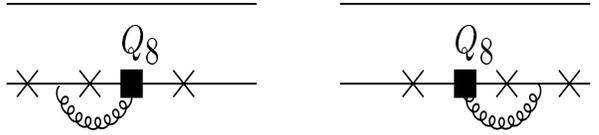}}
   \vspace*{-1cm}
\caption{${\cal O}(\alpha_s)$ contribution to the hard-scattering 
kernels $T^I_{8}$ from chromomagnetic penguin operator $Q_8$.
\label{fig:q8t1}}
\end{figure}

\begin{figure}[t]
   \epsfysize=2cm
   \epsfxsize=8cm
   \centerline{\epsffile{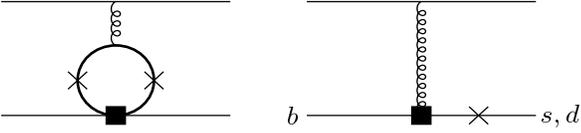}}
   \vspace*{-1cm}
\caption{${\cal O}(\alpha_s)$ contribution to the hard-scattering 
kernels $T^{II}_i$ from four-quark operators $Q_i$ (left) and
from $Q_8$. \label{fig:qit2}}
\end{figure}

The matrix elements of the four-quark operators $Q_i$ (and of $Q_8$)
start contributing at ${\cal O}(\alpha_s)$. In this case the factorization
formula becomes nontrivial. The diagrams for the hard-scattering kernels
$T^I_{i}$ are shown in Fig. \ref{fig:qit1} 
for $Q_1,\ldots, Q_6$ and in Fig. \ref{fig:q8t1}
for $Q_8$. The non-vanishing contributions to
$T^{II}_i$ are shown in Fig.~\ref{fig:qit2}. 

The first negative moment of the $B$-meson LCDA $\Phi_{B}(\xi)$, 
which will be needed below,
can be pa\-ra\-me\-trized by a quan\-ti\-ty 
$\lambda_B={\cal O}(\Lambda_{QCD})$, i.e.
$\int^1_0 d\xi \Phi_{B}(\xi)/\xi=m_B/\lambda_B$.

\begin{figure}[t]
   \epsfysize=2cm
   \epsfxsize=4cm
   \centerline{\epsffile{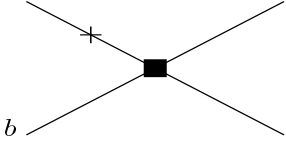}}
   \vspace*{-1cm}
\caption{Annihilation contribution to $\bar B\to V\gamma$ decay.
The dominant mechanism is the radiation of the photon from
the light quark in the $B$ meson, as shown. This amplitude is
suppressed by one power of $\Lambda_{QCD}/m_b$, but it is still
calculable in QCD factorization. Radiation of the photon from the
remaining three quark lines is suppressed by $(\Lambda_{QCD}/m_b)^2$
for operators $Q_{1,2}$. \label{fig:ann}}
\end{figure}

There are further mechanisms that can in principle contribute
to the $\bar B\to V\gamma$ amplitude.
One possibility is weak annihilation (Fig. \ref{fig:ann}),
which is suppressed by $\Lambda_{QCD}/m_b$.
Still the dominant annihilation amplitude
can be computed within QCD factorization because the
colour-transparency argument applies to the emitted, highly energetic
vector meson in the heavy-quark limit \cite{BBNS}.

\section{$B\to K^*\gamma$}\label{sec:BKgam}

In the case of $B\to K^*\gamma$ the component of the
Hamiltonian (\ref{heff}) proportional to $\lambda_u$ is strongly
CKM suppressed ($|\lambda_u/\lambda_c|\approx 0.02$) and has
only a minor impact on the decay rate. It is essentially
negligible, but will be included for completeness.
Here we shall 
neglect the contribution from the QCD penguin
operators $Q_3,\ldots, Q_6$, which enter at ${\cal O}(\alpha_s)$ and
are further suppressed by very small Wilson coefficients.
(The complete results are given in \cite{SWB}.)
We note that to ${\cal O}(\alpha_s)$ the matrix element of $Q_2$
is zero because of its colour structure.
The result for the diagrams in Figs. \ref{fig:qit1} and \ref{fig:q8t1},
which enter the hard-scattering kernels $T^I_{1}$, $T^I_{8}$,
can be infered from \cite{GHW}. In these papers the diagrams were
computed to obtain the matrix elements for the inclusive mode
$b\to s\gamma$ at next-to-leading order. In this context
Figs. \ref{fig:qit1} and \ref{fig:q8t1} represented the virtual
corrections to the inclusive matrix elements of $Q_1$ and $Q_8$. 
In our case they determine the kernels $T^I_{1}$ and $T^I_{8}$.
The results from \cite{GHW} imply
\begin{equation}\label{q1me1}
\langle Q_{1,8}\rangle^I=\langle Q_7\rangle 
\frac{\alpha_s C_F}{4\pi} G_{1,8}
\end{equation}
where $G_1=G_1(s_c)$ and $G_8$ can be found in \cite{BB}
$(s_c\equiv m^2_c/m^2_b)$.

\begin{figure}[t]
   \epsfysize=6cm
   \epsfxsize=7cm
   \centerline{\epsffile{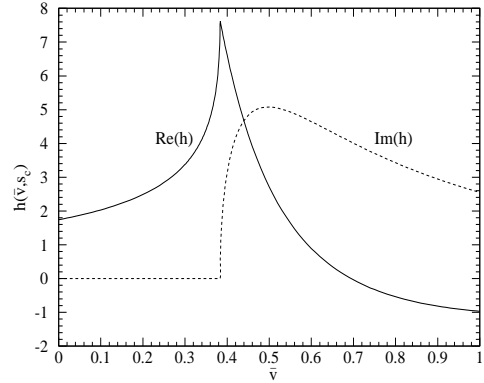}}
   \vspace*{-1cm}
\caption{The hard-scattering kernel $h(\bar v,s_c)$ as a function
of $\bar v$. \label{fig:hcvbar}}
\end{figure}

We now turn to the mechanism where the spectator participates
in the hard scattering.
The first diagram in Fig. \ref{fig:qit2} yields
\begin{equation}\label{q1me2}
\langle Q_1\rangle^{II}=\langle Q_7\rangle 
\frac{\alpha_s(\mu_h) C_F}{4\pi} H_1(s_c)
\end{equation}
with ($\bar v\equiv 1-v$)
\begin{eqnarray}\label{h1s}
&&H_1(s)=-\frac{2\pi^2}{3 N}\frac{f_B f^\perp_V}{F_V m^2_B}
\int^1_0 d\xi\frac{\Phi_{B1}(\xi)}{\xi}\cdot \nonumber \\
&&\ \ \ \ \ \ \ \cdot\int^1_0 dv\, h(\bar v,s) \Phi_\perp(v)
\end{eqnarray}
The function $h(\bar v,s_c)$ is shown in
Fig. \ref{fig:hcvbar}.
The correction to $\langle Q_8\rangle$ from the hard spectator
interaction comes from the second diagram in Fig. \ref{fig:qit2}:
\begin{equation}\label{q8me2}
\langle Q_8\rangle^{II}=\langle Q_7\rangle 
\frac{\alpha_s(\mu_h) C_F}{4\pi} H_8
\end{equation}
Combining these results gives
\begin{eqnarray}\label{abkgam}
&&A(\bar B\to K^*\gamma)= \\
&&\ \ \frac{G_F}{\sqrt{2}}
\left[\sum_p \lambda_p^{(s)}\, a^p_7(K^*\gamma)\right] 
\langle K^*\gamma|Q_7|\bar B\rangle \nonumber
\end{eqnarray}
where, at NLO
\begin{eqnarray}\label{a7vgam}
&&a^p_7(V\gamma) = C_7 + \\
&&\ +\frac{\alpha_s(\mu) C_F}{4\pi}
\left( C_1(\mu) G_1(s_p)+ C_8(\mu) G_8\right) \nonumber \\  
&&\ + \frac{\alpha_s(\mu_h) C_F}{4\pi} 
  \left( C_1(\mu_h) H_1(s_p)+ C_8(\mu_h) H_8\right) \nonumber
\end{eqnarray}

\section{$B\to\rho\gamma$}\label{sec:Brhogam}

For the decay $\bar B\to\rho\gamma$ both sectors of the effective 
Hamiltonian have the same order of magnitude. 
The amplitude is similar to (\ref{abkgam}) with obvious replacements.
The rate for the CP-conjugated mode $B\to\rho\gamma$ 
is obtained with $\lambda_p^{(d)} \to \lambda_p^{(d)*}$. 
We may then consider the CP asymmetry
\begin{equation}\label{acpbrgdef}
{\cal A}_{CP}(\rho\gamma)=
\frac{\Gamma(B\to\rho\gamma)-\Gamma(\bar B\to\rho\gamma)}{
      \Gamma(B\to\rho\gamma)+\Gamma(\bar B\to\rho\gamma)}
\end{equation}
A non-vanishing CP asymmetry appears at ${\cal O}(\alpha_s)$ only. 

We next comment on the issue of power corrections.
The annihilation effect from operator 
$Q_{1}$ gives a numerically important power correction, because
it receives an enhancement of 
$|C_1/C_7|\sim 3$. This leads to a $30\%$ correction in the
amplitude of the charged mode $B^-\to\rho^-\gamma$.
A general discussion of isospin-breaking power corrections
was given in \cite{KN}, where a sizeable effect of $11\%$
from penguin annihilation related to $Q_6$ was identified.
This contribution is still calculable, while other
terms of the same order are much smaller numerically.

Power corrections can also come from the loops
with up- and  charm quarks, whose leading-power
contributions were computed in (\ref{q1me2}).
These power corrections correspond to the region of
integration where the gluon becomes soft, that is 
$\bar v={\cal O}(\Lambda_{QCD}/m_b)$. Their contribution
is nonperturbative and cannot be calculated in the hard-scattering
formalism. Nevertheless, the expression in (\ref{q1me2}) can be used
to read off the scaling behaviour of these power corrections
in the heavy-quark limit. For the charm loop the kernel approaches
a constant $\sim m^2_b/m^2_c$ in the endpoint region.
Taking into account the linear
endpoint suppression of the wave function $\Phi_\perp$, the integral in 
(\ref{q1me2}) over the region $\bar v\sim \Lambda_{QCD}/m_b$ 
thus contributes a term of order 
$(\Lambda_{QCD}/m_b)^2\times (m_b/m_c)^2 = (\Lambda_{QCD}/m_c)^2$.
That is, we recover the power behaviour of soft
contributions in the charm sector first pointed out in \cite{VOL}.
This was discussed for the inclusive decay $b\to s\gamma$ in
\cite{VOL,BIR} and for the exclusive mode $B\to K^*\gamma$ in
\cite{KRSW}. Numerically this correction is very small ($\sim 3\%$
in the decay rate).
A similar consideration applies to the up-quark sector. In this case
the endpoint behaviour of the kernel is singular $\sim 1/\bar v$,
which now leads to a linear power suppression of the form
$\Lambda_{QCD}/m_b$. This coincides with the scaling
behaviour derived in \cite{BIR} in the context of the
inclusive process.

\section{PHENOMENOLOGY}\label{sec:phen}

In this section we present some numerical results based on the 
QCD analysis at NLO. The input parameters used can be found in \cite{BB}.

We note a sizable enhancement of the leading order value, dominated by the 
$T^I$-type correction. This feature was already observed in the context of 
the inclusive case in \cite{GHW}. A complex phase is generated at NLO, 
where the $T^I$-corrections and the hard-spectator interactions ($T^{II}$) 
yield comparable effects.

\begin{figure}[t]
\epsfig{figure=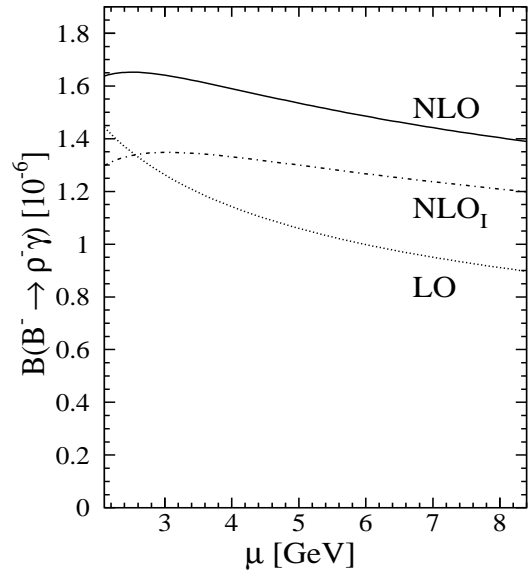,width=7.2cm,height=8.0cm}
\vspace*{-1cm}
\caption{Dependence of $B(B^- \to \rho^- \gamma)$ 
on the renormalization scale $\mu$ at leading (LO) and
next-to-leading order (NLO is the full result,  
$\rm{NLO}_{\rm I}$ the result without hard-spectator
corrections).}
\label{fig:bkrhomu}
\end{figure}

The net enhancement of $a_7$ at NLO leads to a corresponding enhancement 
of the branching ratios, for fixed value of the form factor. This is 
illustrated in Fig. \ref{fig:bkrhomu}, where we show the residual scale 
dependence for $B(B^-\to\rho^-\gamma)$ 
at leading and next-to-leading order.
The uncertainty of the branching fractions is 
currently dominated by the form factors $F_{K^*}$, $F_\rho$. 
Our estimates are (in comparison with the experimental results
in brackets)
$B(\bar{B}\to \bar{K}^{*0}\gamma)/10^{-5}=7.1\pm 2.5$  
$(4.21\pm 0.29 \cite{BABAR1})$
and $B(B^-\to\rho^-\gamma)/10^{-6}=1.6\pm 0.6$ 
$(< 2.3 \cite{BABAR2})$.
Taking the sizable uncertainties into 
account, the results for $B\to K^*\gamma$ are compatible with the 
experimental measurements, 
even though the central theoretical values appear to be somewhat high.
$B(B\to\rho\gamma)$ is a sensitive
measure of CKM quantities such as the angle $\gamma$.
The CP asymmetry ${\cal A}_{CP}(\rho\gamma)$ is of order $10\%$.

\section{$B\to\gamma\gamma$}

The decays $B_{s,d}\to\gamma\gamma$ have recently been
analyzed using QCD factorization based on the heavy-quark limit
\cite{BB2}.
The dominant effect arises from the magnetic-moment type transition
$b\to s(d)\gamma$ where an additional photon is emitted
from the light quark (one-particle reducible diagram). The contributions
from one-particle irreducible diagrams (both photons emitted
from up- or charm-quark loops) are power suppressed,
but still calculable. The dominant effect is very sensitive 
to the $B$ meson parameter $\lambda_B$ defined in sect. 2.

\section{CONCLUSIONS}\label{sec:concl}

In this talk we have discussed a systematic,
model-independent framework for the exclusive radiative
decays $B\to V\gamma$ based on the heavy-quark limit.
This enabled the consistent computation of the decay amplitudes 
at next-to-leading order in QCD.
An important conceptual aspect of this analysis is the interpretation
of loop contributions with charm and up quarks, which come from
leading operators in the effective weak Hamiltonian.
We have argued that these effects are calculable in terms of
perturbative hard-scattering functions and universal meson
light-cone distribution amplitudes. They are ${\cal O}(\alpha_s)$
corrections, but are leading power contributions in the
framework of QCD factorization. This picture is in contrast to the
common notion that considers charm and up-quark loop effects as
generic, uncalculable long-distance contributions.
Non-factorizable long-distance corrections may still exist, but
they are power-suppressed.
The improved theoretical understanding of $B\to V\gamma$ decays
streng\-thens the motivation for still more detailed
experimental investigations, which will contribute
significantly to our knowledge of the flavour sector.

\section*{Acknowledgements}
I thank Stefan Bosch for a most enjoyable collaboration
on the subjects of this talk.

\end{document}